# Phenotype-based and Self-learning Inter-individual Sleep Apnea Screening with a Level IV Monitoring System

Hau-Tieng Wu[1#], Jhao-Cheng Wu[2#], Po-Chiun Huang[2], Ting-Yu Lin[3], Tsai-Yu Wang[3], Yuan-Hao Huang[4*], Yu-Lun Lo[3*&]

1: Department of Mathematics and Department of Statistical Science, Duke University, Durham, NC, USA

2: Department of Electrical engineering, National Tsing-Hua University, Taiwan.

3: Department of Thoracic Medicine, Chang Gung Memorial Hospital, Chang Gung University, School of Medicine, Taipei, Taiwan

4: Department of Electrical engineering and Institute of Communications Engineering, National Tsing-Hua University, Taiwan

#: Dr Wu & Mr. Wu contributed equally to this paper

*: Drs. Huang & Lo contributed equally to this paper

&: Correspondence (email: loyulun@hotmail.com)

**Conflict of interests**: There is no conflict of interests.

**Author contributorship**: HT Wu, YH Huang, TY Lin, TY Wang and YL Lo design the experiment, carry out the study, and prepare the manuscript. HT Wu, JC Wu and YH Huang carried out the data analysis. PC Huang handles the hardware preparation and data collection. All authors contribute to the final manuscript proofreading.

## Abstract

Purpose: We propose a phenotype-based artificial intelligence system that can self-learn and is accurate for screening purposes, and test it on a Level IV monitoring system. Methods: Based on the physiological knowledge, we hypothesize that the phenotype information will allow us to find subjects from a well-annotated database that share similar sleep apnea patterns. Therefore, for a new-arriving subject, we can establish a prediction model from the existing database that is adaptive to the subject. We test the proposed algorithm on a database consisting of 62 subjects with the signals recorded from a Level IV wearable device measuring the thoracic and abdominal movements and the SpO2. Results: With the leave-one cross validation, the accuracy of the proposed algorithm to screen subjects with an apnea-hypopnea index greater or equal to 15 is 93.6%, the positive likelihood ratio is 6.8, and the negative likelihood ratio is 0.03. Conclusion: The results confirm the hypothesis and show that the proposed algorithm has great potential to screen patients with SAS.

## Introduction

Sleep apnea syndrome (SAS) is a common sleep disorder that affects approximately 14% of adult men and 5% of adult women (Peppard et al., 2013). An even higher prevalence is reported in Swiss population (Heinzer et al., 2015) that the prevalence of moderate-to-severe SBD was 23.4% in women and 49.7% in men. SAS has been known to be related to different diseases, or even public tragedies (Canessa et al., 2011; Golbidi et al., 2012; Leger et al., 2012; Yaggi et al., 2005). Although SAS has received considerable attention, most patients with SAS are not aware of it and are untreated (Gibson, 2004; Young et al., 2009). Therefore, a screening tool, better designed for home screening, is urgently needed. This tool should be easy to install at home, cheap, comfortable and not interfere sleep. Many sensors have been explored for this purpose, including those equipped in different items (Al-Mardini et al., 2014; Koyama et al., 2003), for example, the electrocardiogram (ECG) signal, oximeter signal, sound, nasal airflow measurement, respiration effort measurement, oximeter, and accelerometer. In addition to developing an easy-to-install, inexpensive, and accurate screening monitor, researchers have proposed several artificial intelligence (AI) systems for automatic annotation of the collected signal with high accuracy and, therefore, achieve the screening purpose (Alvarez-Estevez and Moret-Bonillo, 2015). However, the inevitable inter-individual variability issue is less considered in these computer-assisted screening techniques.

In addition to accurately annotating collected signals, an AI system should contain a *self-learning* ability like a sleep expert; that is, we are looking for a system that can perform better when there are more cases with good annotations. In practice, to make a diagnosis on a new-arriving patient, physicians automatically handle the inter-individual variability by taking various phenotypes into account, mainly based on his accumulated practicing experience. In this work, we take this wisdom into account, and propose a phenotype-based self-learning AI system for SAS screening. We test the proposed algorithm on a Level IV screening system (Ferber et al., 1994; Collop et al., 2007), which contains a pulse oximeter for SpO2 detection and two tri-axial accelerator (TAA) sensors for thoracic and abdominal movement, which are surrogates of the respiratory signal. For a new-arriving patient, based on the designed phenotype metric based on the clinical phenotypes (body mass index (BMI), age, gender and comorbidity history) and SpO2 and respiratory signals, a prediction model is established from those subjects in the available annotated database that are most similar to the new-arriving subject.

To evaluate the performance of the proposed phenotype-based self-learning AI, we compare the automatic annotations with expert labeled sleep records.

## Material

The study was performed with at least six hours of sleep recording time to confirm the presence or absence of OSA from the clinical subjects suspected of sleep apnea at the sleep center in Chang Gung Memorial Hospital (CGMH), Linkou, Taoyuan, Taiwan. The Institutional Review Board of CGMH approved the study protocol (No. 101-4968A3). The SpO2 is recorded by the Alice 5 data acquisition system (Philips Respironics, Murrysville, PA) sampled at 1 Hz. The thoracic and abdominal movements are simultaneously recorded at 226Hz with the 8 bits resolution (The TAA sensors are ADXL335 (Analog device), and the micro-controller AT328P (Atmel) executes the data acquisition flow and then transfer the data to a server wirelessly by a Bluetooth channel), and the signals are synchronized with the SpO2 signal. We also collected the questionnaire from the subject, including age, gender, height, weight, medical history, and drug history. An apnea event (OSA or CSA) is identified when the airflow breathing amplitude decreases more than 90% for a duration ranging from 10 to 120 seconds, whereas a hypopnea event is identified when the airflow breathing amplitude decreases over 30% of the pre-event baseline with $\geq$ 3% oxygen desaturation or with an arousal (AASM 2012).

## Method

We designed a phenotype-base metric to determine the *similarity* between subjects. The flowchart of the algorithm is illustrated in Figure 1. We consider the commonly available phenotype information for each subject, including gender, age, and body-mass index (BMI) that are closely related to the sleep apnea pattern and severity (Liu et al., 2017). The similarity is designed based on the physician's clinical experience; that is, the closer the age and BMI are, the more similar two subjects are, and the similarity between two subjects with the same gender are weighted more. We call the designed similarity the *phenotype metric*. In clinics, the gender, BMI and age are not the only considered parameters for the SAS. We further take the comorbidity of hypertension, diabetes, and hypothyroidism into account to better determine the similarity between subjects, which is called the *correction distance*. Two subjects with the same comorbidity are more similar. Following the clinical practice, if we want to find the K most similar subjects of the new-arriving subject, we first determine K + K' most similar subjects that are related to the phenotype metric, and remove the K' subjects with the largest correction distance. If there are less than K' subjects that have the correction distance greater than 0, we remove subjects with the largest phenotype distance to determine the K most similar subjects. We call this a *modified K nearest neighbor (KNN) scheme*. The detailed description of the metric design can be found in the online supplementary material.

For each subject, we extract two sets of features – the apnea-related features and the desaturation features. These features are extracted from 10 second long segmented signals, with a 9.5 second overlap. For each segment of the recorded thoracic and abdominal movement signal, we extract the amplitude, frequency, and paradoxical movement as the apnea-related features that are introduced in (Lin et al., 2016). For each segment of the SpO2 signal, the minimum, maximum, median, mean, variance of the first derivative, and difference between the median and minimum over a sliding 20-second window are selected as the desaturation features.

With the *modified KNN scheme* and selected features, the proposed phenotype-based self-learning AI system is carried out upon the available database with annotations provided by the sleep experts in the following way. For the new-arriving subject called Z, we find K most similar subjects by the modified

KNN scheme. The kernel support vector machine (SVM) (Khandoker et al., 2009) based on the standard radial based function is applied to establish a prediction model from the features extracted from those K most similar subjects. The established SVM classifier, combined with the paradoxical movement feature, is applied to design a state machine (Lin et al., 2016). The established state machine is applied to predict the sleep apnea stage of the new-arriving subject Z from the recorded SpO2 and thoracic and abdominal movement signals. Finally, for all epochs classified as normal, if there is a desaturation determined by the desaturation features, that epoch is corrected to an apnea event. With the final whole night sleep apnea annotation, we could estimate the AHI and, therefore, the severity of SAS. Since the sleep and awake information is not available, the AHI is estimated by the respiratory event index (REI). For reproducibility purposes, the detailed description of the feature extraction and state machine is shown in the online supplementary material.

## Assessment

To evaluate the proposed phenotype-based inter-individual classification performance, we apply the leave-one-out cross validation (LOOCV). Each subject was selected for the testing group and the remaining subjects were used for training. We up-sample the training dataset by uniformly duplicating the cases in the smaller subgroup to alleviate the imbalanced case numbers. For the selected subject, we find K most similar subjects from the up-sampled training dataset, and establish the prediction model. We then apply the prediction model on the selected subject. The results of all subjects were averaged to obtain inter-individual testing results. Note that this LOOCV mimics the new-arriving subject in the real world scenario.

We report two aspects of the performance – the event identification and the severity prediction. An accurate apnea events detection algorithm should identify those apnea events in the right location. It means that an identified apnea event should overlap an annotated apnea event provided by the sleep expert. Without this information, although the estimated events might still provide a reasonable AHI, the predicted events could not provide more information. A detected event is classified as true positive if it overlaps with an annotated event; if there is an annotated event but no event is detected, the detection result is classified as a false negative. We report the positive predictive value (PPV), or the precision, and the F1 score, which is the harmonic mean of recall (sensitivity) and PPV. We report the summary statistics by median±median absolute deviation (MAD).

To report the performance of the severity prediction, including normal, mild, moderate, and severe, we report a 4-by-4 confusion matrix M. A summarized overall accuracy (AC), and sensitivities and PPV for each group are reported. For the purpose of screening subjects with severe sleep apnea, we divide subjects into two groups, one with subjects having an AHI greater than or equal to 15 and one less than 15, and report not only the sensitivity, specificity and AC, but also the positive likelihood ratio (LR+) and the negative likelihood ratio (LR-). The whole analysis is carried out in Matlab R2014b with the provided SVM module.

## Results

We enrolled 63 adult snoring subjects over 20-year-old from the outpatient clinic continuously from Sep. 2015 to Aug. 2016. The questionnaire of one subject was missing, so we excluded this case from the study. Two sleep experts identified, marked, and classified the overnight sleep records into two categories, normal (NOR) and apnea (APN), including obstructive sleep apnea, central sleep apnea, mixed-type sleep apnea, and hypopnea. Among 62 subjects, there are 49 males and 13 females, 10 normal subjects, and 11, 4 and 37 subjects with mild, moderate and severe SAS respectively. The age is 34.8±16.3, 38.6±15.5, 49.8±13.1, and 52.3±13.8 for the normal, mild, moderate and severe group, respectively. More demographic detail information, including gender, AHI, BMI, age and sleep recording time is summarized in Table 1.

The event-by-event detection results are shown in Table 2. Overall, the PPV is 0.67±0.23 and the F1 is 0.7±0.22, and the algorithm is more accurate for subjects with moderate and severe SAS. For subjects with AHI less than 15, the PPV is 0.24±0.24 and the F1 is 0.27±0.18; for subjects with AHI greater than 15, the PPV is 0.77±0.12 and the F1 is 0.77±0.11.

Table 3 shows the confusion matrix of the severity prediction results, based on the event-by-event prediction result, where the overall accuracy is 71% for 4 groups. The sensitivities are 60%, 63.6%, 75%, and 75.7% for the normal, mild, moderate and severe groups respectively; the PPVs are 85.7%, 58.3%, 20% and 100% respectively. If we take AHI 15 as the cutoff to determine if a subject has an urgent treatment need for his/her SAS, the sensitivity is 97.6%, the specificity is 85.7%, and the accuracy is 93.6%, with LR+ 6.8 and LR- 0.03.

# Discussion

We propose a phenotype-based inter-individual SAS screening algorithm based on the proposed phenotype metric. For each new-arriving subject, we establish a predictor from the K most similar subjects in an existing database with annotations. The predictor is clearly adaptive to the new-arriving subject. We evaluate the algorithm on a database with a Level IV monitoring system equipped with TAA sensors capturing the thoracic and abdominal movements and an oximeter capturing the SpO2 and report the results. If we are concerned with classifying subjects into normal, mild, moderate or severe subgroups, the proposed algorithm achieves 71% accuracy. If we are concerned with screening subjects with an urgent need for SAS treatment, which are those subjects with AHI greater than or equal to 15, the overall accuracy achieves 93.6% and LR- is as low as 0.03. The low LR- means that the proposed algorithm could efficiently rule out the possibility of moderate or severe SAS, and, therefore, accurately screen those patients with moderate or severe SAS based on the Level IV portable device. On the other hand, since LR+ is 6.8, which only moderately increases the post-test probability of disease, the proposed algorithm is not suitable for diagnostic purposes.

For the event-by-event detection result, overall the proposed algorithm could achieve PPV= 0.67±0.23 and F1=0.7±0.22, and the prediction accuracy is better when AHI is higher. When a subject is normal or has a mild SAS, the event-by-event prediction is not good. This is because, based on the metric, we find neighbors that have similar sleep apnea behavior, and there are limited sleep apnea events to train an efficient SVM classifier. On the other hand, we have a higher accuracy for the group with more severe patients, since more apnea events are available. The event-by-event detection is important for several clinical applications. For example, we need a real-time and accurate event-by-event predictor to establish an adaptive continuous positive airway pressure (CPAP) machine. Since the proposed algorithm performs better for subjects with AHI greater than 15, we could expect its clinical potential to improve the CPAP compliance of those patients who urgently need a treatment.

Note that we determine the SAS severity by evaluating REI from the Level IV equipment. While it is widely accepted that REI determined from the PSG underestimates AHI, our estimated REI is different from that determined from the PSG. Based on the event-by-event detection analysis, where we faithfully take any false positive detection during the awake stage into account, the underestimation of the estimated REI is alleviated. By combining the above reported results, we could confirm the potential of the proposed algorithm as a screening tool.

The main obstacle toward a self-evolving capability of a system is the inevitable variation among individuals, and our solution is encoding the physicians' decision making process and clinical experience into the AI system. To the best of our knowledge, this phenotype-based approach to handle inter-individual variability was never considered in the existing computer-assisted SAS screening techniques. See, for example, (Álvarez et al., 2017; Shokoueinejad et al., 2017), and the literature cited therein, for the recent systematic review of computer-assisted SAS screening techniques. The SAS is a reflection of the complicated interaction between different underlying physiological systems and the environment. The interaction varies from subject to subject, so the signals we collect also vary from subject to subject. Due to this inter-individual variation, the model established from the *whole* database might be blurred. For example, two subjects of different genders might express their SAS patterns differently in the collected signals, and the model established from males might not accurately predict the SAS severity of a female. This "blurring effect" deteriorates the performance of the AI system, and the larger the database is, the more severe the "blurring effect" caused by the inter-individual variability will be. As a result, no matter how large the database is, the accumulated knowledge is limited. Therefore, finding a way to "compare" individuals and to select a suitable subset from the database to establish the prediction model for the new-arriving subject becomes a critical problem. This problem could be understood as the "metric design" problem in the machine learning field, which is the main component of our work. With the designed phenotype metric, the "blurring effect" could be alleviated and hence the self-evolving system is possible. The metric is designed based on the physician's experience and interpretation in order to alleviate the influence of the inter-individual variability. This fact has been shown in the reported result -- the modified KNN scheme helps us to select subjects sharing similar features, which improves the prediction accuracy. In general, if we have a more intact electrical health recording system, more phenotype information can be taken into account to design the desired metric.

According to 2007 JCSM guideline (Collop et al., 2007), the equipment we consider to prove the concept of the proposed self-learning system technically does not fall in the category of Type 4 devices while it does not fall in the category of Type 3 neither, since we have less than 4 channels. However, since the sensors we consider collect two types of information – respiratory effort and oxygen saturation, it is closer to a Type 4 device in 2007 JCSM guideline.

There are several technical details regarding the algorithm that must be discussed. We emphasize that although we could run a greedy optimization to determine the optimal weight for each phenotype parameter for the phenotype distance, we do not do it to avoid over-fitting, due to the case number limitation. Second, while the selected features and SVM overall performs well, it is widely accepted that when the database is large, the multi-layer neural network might perform better. In the future large scale study, we could consider designing a multi-layer neural network to replace SVM. Third, to purely study the potential of the algorithm, we do not consider the signal quality effect. All recorded signals are taken into account for the analysis. For a practical application, we could consider distinguishing between signals with high and low qualities. Designing a signal quality index for the TAA signal is a research topic of its own interest, but is out of the scope of this paper. It will be carried out in the future research, and in general this could improve the result.

Despite the strength of the proposed algorithm, we acknowledge several limitations. First, the case number is small and there is a subgroup (female with moderate SAS) that is empty. Based on this preliminary study, a large scale prospective study is needed. With a larger database, we could further take more physicians' wisdom into account. For example, it is known that menopause females have an increases prevalence of SAS. The metric design should take this into account. Second, while the proposed event-by-event detection algorithm has the potential for the clinical application, like improving the CPAP machine, its accuracy could be further improved. Third, the subjects all slept in the lab (not at home) for only one night. Therefore, the first night effect (Tamaki et al., 2016) is inevitable. A prospective study designed for a home care scenario is also needed.

## Conclusion

We confirm that the proposed novel phenotype-based inter-individual SAS prediction algorithm based on a Level IV monitoring system has potential as a self-learning AI system for homecare screening.

## Funding

No funding was received for this research.

## Ethical approval

All procedures performed in studies involving human participants were in accordance with the ethical standards of the institutional and/or national research committee (name the institution/committee) and with the 1964 Helsinki declaration and its later amendments or comparable ethical standards.

## Informed consent

Informed consent was obtained from all individual participants included in the study.

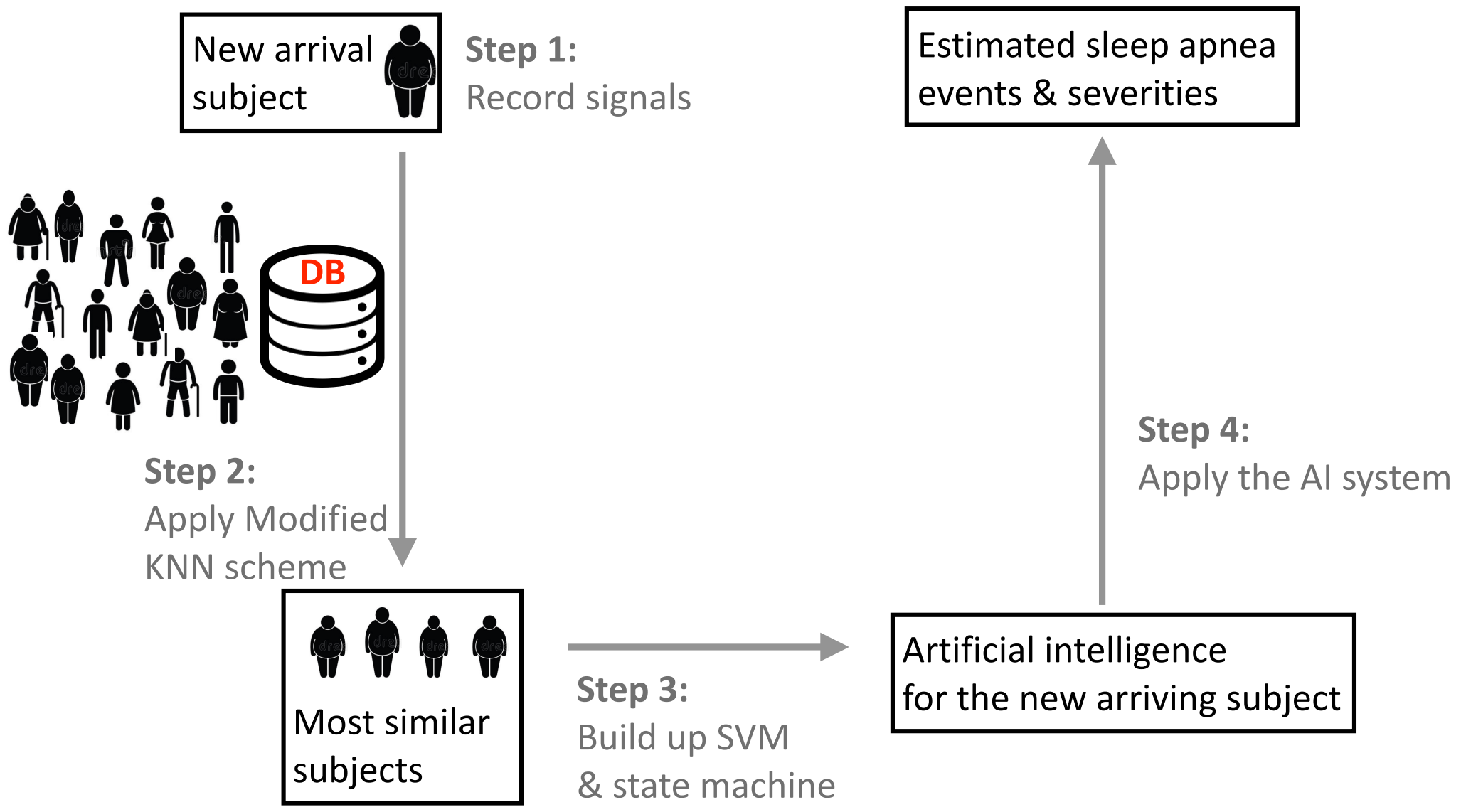

**Figure 1:** The flowchart of the proposed phenotype-based self-learning artificial intelligence system, with the prototype of the sleep apnea screening instrument. In the clinical testing environment, a full-featured PSG system is also attached to record the ground truth. Two tri-axial accelerators are attached on the backside of piezo belts. The oximeter probe is applied to the forefinger of right hand, which is not shown in this photo. DB indicates the existing database with experts' annotations and phenotyp information.

| | gender (# of sub.) | AHI (#/hour) | BMI (kg/m$^2$) | Age (y/o) | Sleep Time (hour) | # of CSA | # of MSA | # of OSA | # of HYP |
|---|---|---|---|---|---|---|---|---|---|
| **Normal** | All (10)<br>Male (4)<br>Female (6) | 2.2 ± 1.4<br>2.4 ± 1.0<br>2.1 ± 1.8 | 22.4 ± 2.8<br>22.1 ± 1.4<br>22.6 ± 3.5 | 34.8 ± 16.3<br>36.8 ± 17.0<br>33.5 ± 17.2 | 6.3± 0.2<br>6.3 ± 0.2<br>6.3 ± 0.2 | 2.1 ± 2.1<br>1.3 ± 0.5<br>2.7 ± 2.7 | 0.5 ± 0.7<br>0.3 ± 0.5<br>0.7 ± 0.8 | 1.0 ± 2.5<br>0.5 ± 0.6<br>1.3 ± 3.3 | 8.5 ± 6.4<br>10.8 ± 5.3<br>7.0 ± 7.1 |
| **Mild** | All (11)<br>Male (7)<br>Female (4) | 9.9 ± 2.7<br>9.6 ± 3.3<br>10.4 ± 1.4 | 25.0 ± 4.5<br>23.5 ± 3.9<br>27.5 ± 4.8 | 38.6 ± 15.5<br>29.9 ± 8.1<br>53.8 ± 13.8 | 6.3 ± 0.1<br>6.3 ± 0.1<br>6.1 ± 0.1 | 3.1 ± 3.4<br>3.9 ± 4.1<br>1.8 ± 1.5 | 1.8 ± 1.8<br>2.4 ± 2.0<br>0.8 ± 0.5 | 14.9 ± 13.0<br>18.0 ± 15.3<br>9.5 ± 5.5 | 34.4 ± 11.8<br>29.0 ± 7.3<br>43.8 ± 13.1 |
| **Moderate** | All (4)<br>Male (4)<br>Female (0) | 24.9 ± 5.3<br>24.9 ± 5.3<br>– | 27.0 ± 1.6<br>27.0 ± 1.6<br>– | 49.8 ± 13.1<br>49.8 ± 13.1<br>– | 6.4 ± 0.3<br>6.4 ± 0.3<br>– | 5.0 ± 10.0<br>5.0 ± 10.0<br>– | 3.3 ± 5.9<br>3.3 ± 5.9<br>– | 18.3 ± 16.6<br>18.3 ± 16.6<br>– | 96.0 ± 41.1<br>96.0 ± 41.1<br>– |
| **Severe** | All (37)<br>Male (34)<br>Female (3) | 63.8 ± 23.4<br>63.9 ± 23.6<br>62.6 ± 24.9 | 27.8 ± 3.7<br>27.6 ± 3.5<br>29.6 ± 5.9 | 52.3 ± 13.8<br>51.0 ± 13.6<br>67.7 ± 1.2 | 6.3 ± 0.1<br>6.3 ± 0.1<br>6.2 ± 0.1 | 9.1 ± 14.9<br>9.6 ± 15.5<br>4.3 ± 1.2 | 22.3 ± 32.7<br>21.6 ± 32.7<br>30.7 ± 39.4 | 179.6 ± 121.9<br>181.4 ± 124.5 | 103.5 ± 71.7<br>103.5 ± 74.7<br>103.3 ± 23.9 |

**Table 1**: Demographic details of the enrolled 62 subjects.

| | | |
|---|---|---|
| PPV | Normal | 0.10 ± 0.26 |
| | Mild | 0.38 ± 0.19 |
| | Moderate | 0.49 ± 0.07 |
| | Severe | 0.80 ± 0.11 |
| | All | 0.67 ± 0.23 |
| F1 | Normal | 0.17 ± 0.16 |
| | Mild | 0.36 ± 0.16 |
| | Moderate | 0.56 ± 0.07 |
| | Severe | 0.81 ± 0.10 |
| | All | 0.70 ± 0.22 |

**Table 2**: Results of the event-by-event prediction of proposed phenotype-based inter-individual predictor. In the proposed algorithm, K=15 and K'=5. PPV=positive predictive value.

| | | **Expert label** | | | |
|---|---|---|---|---|---|
| | | Normal | Mild | Moderate | Severe |
| **Prediction** | Normal (AHI < 5) | 6 | 1 | 0 | 0 |
| | Mild (5 < AHI ≤ | 4 | 7 | 1 | 0 |
| | Moderate (15 < | 0 | 3 | 3 | 9 |
| | Severe (30 < AHI) | 0 | 0 | 0 | 28 |
| Accuracy | | 70.97% | | | |

**Table 3**: Confusion matrix of the proposed phenotype-based inter-individual prediction algorithm. In the proposed algorithm, K=15 and K'=5.